# Maunakea Spectroscopic Explorer Advancing from Conceptual Design

Kei Szeto[1a], Doug Simons[a], Steven Bauman[a], Alexis Hill[a], Nicolas Flagey[a], Alan McConnachie[b], Shan Mignot[c], Richard Murowinski[a]

[a] CFHT Corporation, 65-1238 Mamalahoa Hwy, Kamuela, Hawaii 96743, USA
[b] National Research Council Canada, Herzberg Astronomy and Astrophysics, 5071 West Saanich Road, Victoria, BC, Canada, V9E 2E7
[c] GEPI, Observatoire de Paris, PSL Research University, CNRS, Univ Paris Diderot, Sorbonne Paris Cité, Place Jules Janssen, 92195 Meudon, France

## ABSTRACT

The Maunakea Spectroscopic Explorer (MSE) project has completed its Conceptual Design Phase. This paper is a status report of the MSE project regarding its technical and programmatic progress. The technical status includes its conceptual design and system performance, and highlights findings and recommendations from the System and various subsystems design reviews. The programmatic status includes the project organization and management plan for the Preliminary Design Phase. In addition, this paper provides the latest information related to the permitting process for Maunakea construction.

## 1. INTRODUCTION

This paper serves as a technical and programmatic status report of the MSE project as it emerges from the CoDP and progresses toward the Preliminary Design Phase (PDP). MSE, as a planned project to replace the existing 3.6-m Canada France Hawaii Telescope with a dedicated wide field highly multiplexed fiber fed spectroscopic facility, was presented in the introductory papers by Murowinski[1] and Szeto[2] as the project began its Conceptual Design Phase (CoDP) in 2016. Murowinski described the project organization, partnership structure and its corresponding work share, and on the permitting considerations as related to the Hawaiian community. Szeto described the observatory baseline design and its system decomposition, multiplexing options for the positioner system and spectrograph systems such that MSE is capable of observing over four thousand science targets simultaneously in two resolution modes, low or moderate and high.

Figure 1 compares the initial observatory configuration with the configuration that emerged from the CoDP. Main differences include the telescope structure, inner pier modifications and the location of the high resolution spectrographs. The large telescope elevation journals are replaced by compact trunnions to improve structural performance over the full range of zenith pointing. The high resolution spectrographs are moved to the lower floor pier lab. The ceiling of the existing upper Coude room is removed to create an expanded space for the central telescope support structure, azimuth cable wrap, lateral restraints and seismic damper system. All of these components are located below the telescope azimuth track.

---

[1] Email: szeto@cfht.hawaii.edu; Telephone: 808-885-3188; Fax 808-885-7288

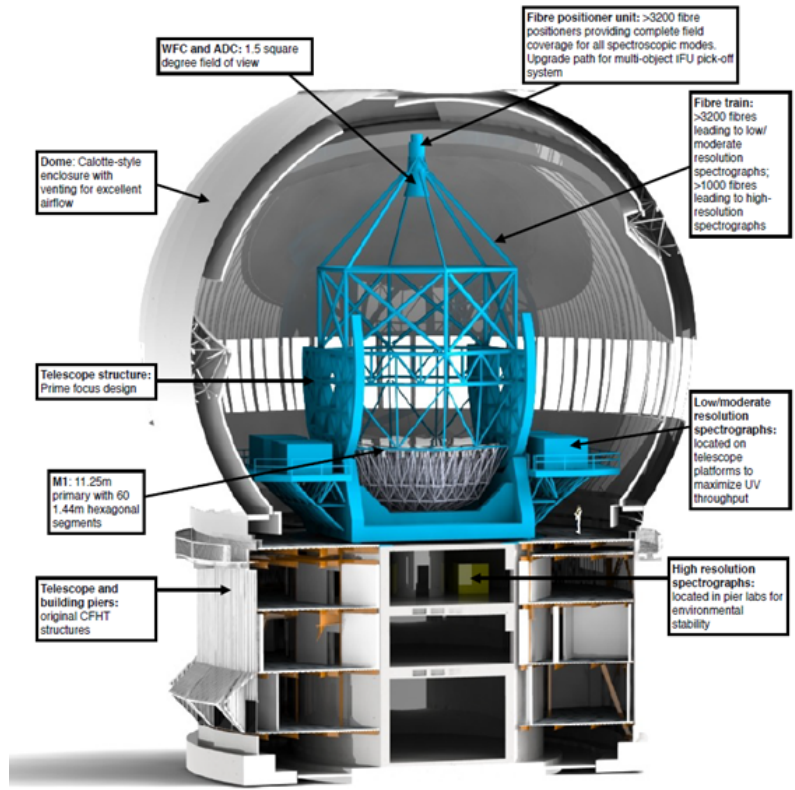

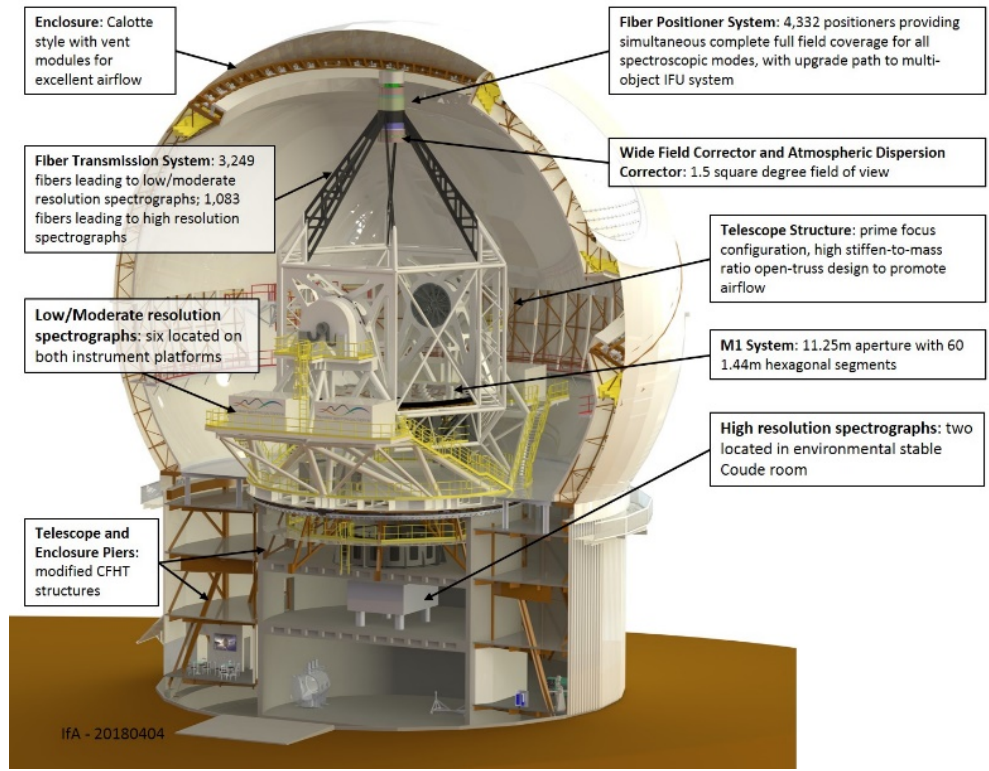

Figure 1 MSE observatory configuration - initial design (top) and conceptual design (bottom)

## 2. DESIGN DESCRIPTION OF THE OVERALL OBSERVATORY SYSTEM

Figure 2 shows the MSE product breakdown structure (PBS) as the project emerges from the CoDP. the current MSE system decomposition is essentially unchanged from the initial observatory configuration.

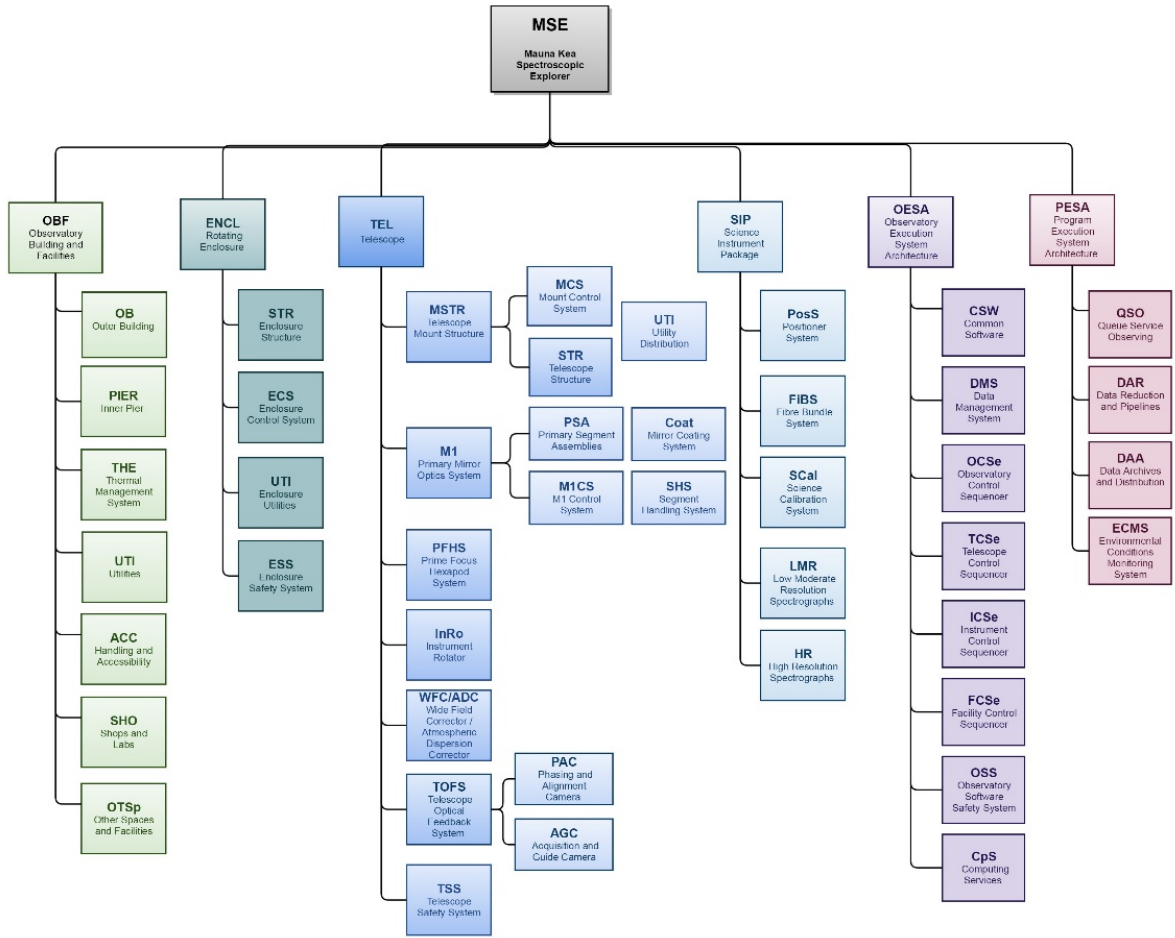

Figure 2 MSE product breakdown structure

The MSE Observatory is organized under six top level products:
- Observatory Building Facilities
- Enclosure
- Telescope
- Science Instrument Package
- Observatory Execution System Architecture
- Program Execution System Architecture

**2.1 Observatory Building Facilities (OBF)**

Structurally, the OBF contains two independent pier structures that support the enclosure and telescope systems. Operationally, the OBF contains the infrastructure to enable science operations, including mechanical and electrical plants, laboratories for coating mirror segments and servicing science instruments, shops for servicing and maintenance of the enclosure and telescope components, and personnel space such as offices, technical library, staff lounge, first aid station, etc.

**2.2 Enclosure (ENCL)**

The ENCL is a Calotte style dome with independently rotating base and cap structures. The base contains ventilation modules, an enclosure mounted crane, a telescope top-end service platform and a fixed shutter structure. The cap contains the aperture opening and rotates on a plane inclined at 30 ° from the horizon, at half of the telescope zenith range, atop the base structure. The combined base and cap rotations enable full sky coverage for the telescope.

The ENCL also includes a hardware-based safety system for protection of personnel and equipment.

**2.3 Telescope (TEL)**

The TEL has an Alt/Az mount[3] and provides structural attachment points for the telescope optics (segmented primary mirror and wide field corrector) and science instrument package. The telescope optics deliver a 1.5 deg$^2$ focal surface at the prime focus station. The telescope top-end contains hexapod and instrument rotator systems. Mechanically, the hexapod carries the wide field corrector (WFC) optical barrel and the instrument rotator, i.e. field de-rotator, which in turn carries the prime focus station. These three subsystems comprise the top end assembly (TEA). The prime focus station contains the positioner system and two camera systems (one for acquisition and guiding, and the other for segment alignment, phasing and warping). Collectively, they are contained in the Telescope Optical Feedback System PBS element.

The TEL distributes the utilities required to operate its mount and top-end components, and the telescope mounted science instrument package. The TEL also includes a hardware-based safety system for protection of personnel and equipment.

**2.4 Science Instrument Package (SIP)**

The SIP contains the components that are directly associated with obtaining science data. At the prime focus station, the positioner system acquires science targets by positioning the inputs of the fibre transmission system which is the light conduit between the focal surface to the spectrograph input slits. Two types of spectrographs are available for (switchable) low or moderate resolution (LMR) and high resolution (HR) observations.

The science calibration system provides the reference arcs and flats that are critical to characterize the end-to-end system in order to achieve the spectrophotometry and sky subtraction precision required.

**2.5 Observatory Execution System Architecture (OESA)**

The OESA contains the computer network with control hardware and software systems that enable science operations to collect and store science data by receiving commands from the Program Execution System Architecture described in the next section. The OESA is designed to support remote observing from the Waimea headquarters and provides additional functionalities to facilitate engineering operations.

The OESA also includes a hardware-based global safety system for protection of personnel and equipment of the entire observatory.

**2.6 Program Execution System Architecture (PESA)**

The PESA is a collection of high-level software modules that provide the following functionalities to facilitate science operations:
- Schedule and direct observations for science and calibration data
- Reduce and analyze science data to provide real-time feedback for scheduling

- Archive and distribute raw and reduced science data along with the associated calibration and environmental information

In addition, PESA contains sensors to monitor the environmental conditions for the purpose of scheduling, grading and monitoring the progress of science observation to ensure the observatory is operating within the safe design environmental limits.

## 3. ENGINEERING STATUS OF THE OBSERVATORY SYSTEM AND ITS SUBSYSTEMS

In 2017, the Project Office (PO) conducted ten formal conceptual design reviews (CoDR's) for eight subsystems by eight independent external review panels, including the three competing fibre positioner systems which were reviewed by a common panel. The common panel was asked to recommend an integrated positioner and metrology design for further engineering development in the next design phase.

Figure 3 shows the subsystems which have undergone the formal conceptual design review process. The subsystems that have been reviewed yet are the OBF, PESA and part of the TEL, namely the M1 Primary Mirror Optics System and Telescope Optical Feedback System (TOFS).

Additional work on these subsystems is needed to adequately retire their technical risks to the overall project. The status of this work can be summarized as follows:

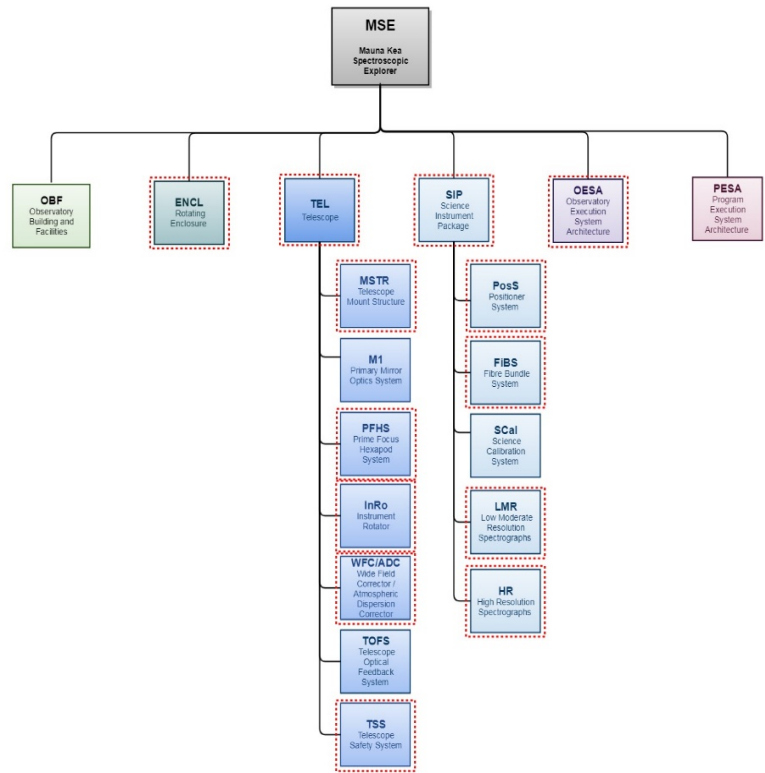

Figure 3 MSE PBS - conceptual design reviewed subsystems are highlighted in red dashed boxes

- An OBF conceptual design review is pending before the end of 2018. Results from on-site geotechnical drilling[4] has confirmed the load bearing capacity of the soil. Engineering studies have validated the structural integrity of the outer building pier and inner pier to support the mass of the new enclosure and telescope, respectively.
- Initial evaluation showed the M1 systems developed for the European Extra-Large Telescope (E-ELT) and Thirty Meter Telescope (TMT), can be adapted for the MSE segmented mirror which has the same size segments.
- Feasibility studies showed the proposed acquisition and guiding camera, and the segment alignment, phasing and warping camera concepts are both viable.
- Functionalities and operational considerations for the PESA system have been outlined in the Operations Concept Document. The corresponding design requirements are listed in the Observatory Requirements Document and the functional and performance requirements of the PESA will be defined for the next design phase.

We determined the technical risks of not having completed conceptual level designs of these subsystems to be acceptable at the conclusion of the CoDP. Nonetheless, we plan to conduct internal reviews of the M1system technical readiness and TOFS before the end of 2018.

**3.1 Highlights from Subsystem Design Reviews – Findings and Recommendations**

Table 1 summaries the major findings and recommendations from the review panels for the eight subsystems reviewed. The positioner system review panel recommended the Australian Astronomical Observatory (AAO) Sphinx[2] fibre positioner[5] system as the primary design to move forward in the PDP and keeping the University of Science and Technology of China (USTC) design as the backup design. Three other review panels recommeded follow-up delta-CoDR's for the subsystems: TEA[6] (hexapod, instrument rotator and WFC optical barrel), fibre transmission system[7] and low/moderate resolution spectrograph[8] as their conceptual designs were judged as either incomplete or presenting high technical risks to the project. The fibre transmission system delta-CoDR is scheduled for July 2018.

| Subsystem and design organization | Major Findings/Recommendations |
|---|---|
| Telescope Structure – IDOM Ingenieria y Consultoria S.A. (Spain) | Design fully meets requirements; the proposed telescope structure reaches preliminary design level in some areas. |
| Enclosure – Dynamic Structures Ltd. (Canada) | Design exceeds mass budget and the review panel provided mass reduction strategies for implementation at early PDP. |
| High Resolution Spectrograph – Nanjing Institute of Astronomical Optics & Technology (China) | Explore optical designs with different dispersion element options and off-axis collimator to accommodate native fibre output at f/2 exit beam. |
| Fibre Positioner Systems and Fibre Position Metrology System – Australian Astronomical Observatory, University of Science and Technology of China and Autonomous University of Madrid, Spain | Three competed designs were reviewed. The review panel had clear preference for the AAO system which is judged to have the lowest programmatic risk. The review panel also acknowledged the USTC system had reached conceptual design level and some areas of the AAO design is at preliminary design level. |
| Fibre Transmission System – Herzberg Astronomy and Astrophysics Research Centre (Canada) | Design not mature enough to proceed to PDP, specifically an end-to-end fibre design concept is missing and a delta-CoDR is recommended. |
| Observatory Execution Software Architecture – Canadian France Hawaii Telescope Corporation (USA) | Design solution of using the CFHT software as a starting point is a good way to reduce cost and risk. However, comprehensive design requirements should be provided as part of the review. |
| Top End Assembly – Division Technique de l'INSU (France) | More work is needed to complete the conceptual design, e.g. missing are the lens mounting design that is based on an optical tolerance budget and service cable wrap concept for the instrument rotator. Therefore, a delta-CoDR is recommended. |
| Low/Moderate Resolution Spectrograph – Centre do Recherche Astrophysique de Lyon (France) | Overall design is assessed to be high risk and a delta-CoDR with alternative lower risk optical designs to reduce number aspheres and allow more space for opto-mechanical packaging is recommended. |

Table 1 Summary findings for design review panels

---

[2] The Sphinx is an evolution of the AAO tilting spine fibre positioner which was first designed and implemented in the FMOS-Echidna instrument for the Subaru Telescope.

## 3.2 Development on Multiplexing Configuration

In his previous paper, Szeto listed six multiplexing configurations proposed for conceptual design studies, including one for each positioner technology (phi-theta and Sphinx). They were selected for engineering development in the CoDP supported by the international design team.

Subsequent to the positioner system review panel recommendation for the Sphinx system, we conducted a detailed down-select analysis between the two positioner and metrology system conceptual designs as presented by AAO and USTC. Included in the analysis was, an evaluation matrix to rank system attributes such as performance, interface requirements, reliability and operational considerations in more than 40 areas organized in nine categories (Table 2). The AAO system scored higher than the USTC system in the evaluation matrix.

| | Criterion | Unit | Priority | Echidna | Phi-Theta UAM | Phi-Theta USTC |
|---|---|---|---|---|---|---|
| t | Item | | 1 to 3* | Normalized score** | | |
| 1. Utilization at system commission | LMR positioner utility factor | % | | | | |
| | HR positioner utility factor | % | | | | |
| 2. Positioner attributes | RMS closed-loop positional accuracy, with metrology system | um | | | | |
| | Max. closed-loop error | um | | | | |
| | Total focus error at max. radius*** of patrol distance | um | | | | |
| | Configuration time, with metrology system | sec | | | | |
| | Resolution, with metrology system | um | | | | |
| | RMS open-loop accuracy | um | | | | |
| | Stability - 30 minutes, RMS | um | | | | |
| | Stability - 30 minutes, max. | um | | | | |
| | FRD - positioner induced | % | | | | |
| | Attenuation loss - positioner induced | % | | | | |
| | Robustness against shock and vibration | g | | | | |
| | Robustness due to collision | n/a | | | | |
| 3. Focal surface | Number of LMR positioners | n/a | | | | |
| | Fill factor - single LMR positioner | % | | | | |
| | Fill factor - more than one LMR positioners | % | | | | |
| | Fill factor - more than two LMR positioners | % | | | | |
| | Minimum distance between two LMR fibres | mm | | | | |
| | Min. circular diameter between three LMR fibres | mm | | | | |
| | Number of HR positioners | n/a | | | | |
| | Fill factor - single HR positioner | % | | | | |
| | Fill factor - more than one HR positioners | % | | | | |
| | Fill factor - more than two HR positioners | % | | | | |
| | Minimum distance between two HR fibres | mm | | | | |
| | Min. circular diameter between three HR fibres | mm | | | | |
| | Reference structure, assign rating between 1 to 10 | n/a | | | | |

Table 2 Screen shot of the positioner and metrology system down-select matrix

In addition, the survey efficiencies between the two systems, which are the product of injection efficiency[3] (IE) and target allocation efficiency, were compared quantitatively. Despite concerns in IE losses in focal ratio degradation (FRD) and defocus from the Sphinx spine tilt, the difference in the system-level IE[9] between the two systems after all losses are accounted is negligible. The point spread function variations due to spine tilt will be corrected by the science calibration system using our prescribed calibration plan[17].

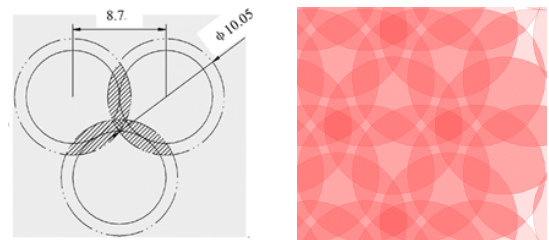

Figure 4 Overlap of positioner patrol area – Sphinx (right) and phi-theta (left)

---

[3] Injection Efficiency is defined as the fraction of flux from an astronomical point source that enters a fibre compared to the flux delivered to the focal surface.

However, the Sphinx system holds a distinct advantage in target allocation efficiency due to its better field coverage as its overlapping poisitioner patrol area is higher than the phi-theta system (Figure 4).

Most importantly, the Sphinx system utilizes two indepentent sets of positioners, 3249 units for LMR amd 1083 units for HR, to provide simultaneous fibre-feed to two sets of spectrographs. This configuration enables parallel LMR and HR observations with full field coverage at each telescope pointing. For the phi-theta system, each of the 3200 positioners carries two fibres, i.e. 3200 fibres for LMR and 3200 fibres for HR. Operationally, this configuration provides either LMR or HR observation exclusively at each telescope pointing. Since the HR spectrographs have a target capacity that is 1/3 of the LMR spectrographs, full field coverage is achieved by using 3200 3-to-1 optical switches to fibre-feed the HR spectrographs.

During the CoDP, we commissioned an optcial switch feasibility study with Durham University to identy candidate commerical products. The best candidate identified was the Luminos CORALIGN$^{TM}$ optical switch which was originally developed for telecommunication applications. Its design provides exceptional mechanical fibre alignment accuracy and repeatability at the micron level; however, its optical performance in terms of FRD has never been tested. Moreover, the feasibility report raised additional uncertainies regarding the output beam point spread function stability, thermal effects (when the switching actuator is activated electrically) and fibre buffer material removal procedures.

Although we believe these concerns are not unsurmountable given sufficient R&D time and resources, the Sphinx system offers simultaneous LMR and HR observations and effectively increase the survey efficiency by 50% over the phi-theta system based on our analysis[10]. We therefore selected the Sphinx system for the PDP, avoiding complications inherent to the optical switching.

### 3.2.1 Selected Multiplexing Configuration for Preliminary Design Phase

The various multiplexing configurations for the fibre subsystems include:
- Sphinx positioner and metrology system design
  - 3249 positioners carrying LMR fibres for the LMR spectrograph system
  - 1083 positioners carrying HR fibres for the HR spectrograph system
- Low and moderate resolution spectrograph design
  - Modular design with a total of 3249 spectra capacity
- High resolution spectrograph design
  - Modular design with a total of 1083 spectra capacity
- Fiber transmission system design
  - Fibre bundles with 3249 fibres feeding the LMR spectrograph system
  - Fibre bundles with 1083 fibres feeding the HR spectrograph system
- Accommodation for future upgrade
  - Modular design such that the positioner system can be replaced by an integral field unit system

### 3.3 System Conceptual Design Review Recommendation

After a series of subsystem reviews in 2017, the Conceptual Design Phase culminated in January 2018 with a System design review to determine if the MSE design will meet the science requirements[11]. Table 3 was presented at the System design review along with our mitigation plan. It shows the compliance of the System conceptual design against the science requirements. It also shows our Risk Exposure (RE) evaluations where the RE is the product of the probability of occurance times the sum total of cost and schedule impacts.

RE = Probability x (Cost Impact + Schedule Impact)

According to our scoring scheme, the minimum RE can range from 2 (3%) to 64 (100%). Currently, the technical Risk Register has a total of eight risks - one LOW score of 8 (13%), six MEDIUM scores ranging from 12 (19%) to 24 (38%) and one HIGH score of 36 (54%).

| Requirements relating to Spectral Resolution: | | Compliance [Y/N/Partial] | Risk Exposure |
|---|---|---|---|
| REQ-SRD-011 | Low spectral resolution | Partial | Medium |
| REQ-SRD-012 | Moderate spectral resolution | Partial | Medium |
| REQ-SRD-013 | High spectral resolution | Y | |
| | | | |
| Requirements relating to the Focal Plane Input: | | | |
| REQ-SRD-021 | Science field of view | Y | |
| REQ-SRD-022 | Multiplexing at low resolution | Partial | |
| REQ-SRD-023 | Multiplexing at moderate resolution | Partial | |
| REQ-SRD-024 | Multiplexing at high resolution | Y | |
| REQ-SRD-025 | Spatially resolved spectra | Y | |
| | | | |
| Requirements relating to Sensitivity | | | |
| REQ-SRD-031 | Spectral coverage at low resolution | Y | |
| REQ-SRD-032 | Spectral coverage at moderate resolution | Y | |
| REQ-SRD-033 | Spectral coverage at high resolution | Y | |
| REQ-SRD-034 | Sensitivity at low resolution | Partial | Medium |
| REQ-SRD-035 | Sensitivity at moderate resolution | Y | |
| REQ-SRD-036 | Sensitivity at high resolution | Partial | High |
| | | | |
| Requirements relating to Calibration | | | |
| REQ-SRD-041 | Velocities at low resolution | TBD | |
| REQ-SRD-042 | Velocities at moderate resolution | TBD | |
| REQ-SRD-043 | Velocities at high resolution | TBD | |
| REQ-SRD-044 | Relative spectrophotometry | Y | |
| REQ-SRD-045 | Sky subtraction, continuum | Y | |
| REQ-SRD-046 | Sky subtraction, emission lines | TBD | Medium |
| | | | |
| Requirements relating to Lifetime Operations | | | |
| REQ-SRD-051 | Accessible sky | Y | |
| REQ-SRD-052 | Observing efficiency | Y | |
| REQ-SRD-053 | Observatory lifetime | Y | |
| | | | |
| Requirements relating to Data Management and Processing | | | |
| REQ-SRD-061 ++ | | | |

Table 3 Tabulated compliance and risk exposure according the science requirements

In Table 3, under the "Compliance" column, Partial means a portion of a multiple-part requirement is met; Partial means open debate whether the requirement is met; and TBD means by design or analysis the requirement will be met with additional work in the next design phase. Under the "Risk Exposure" column, there are five risks, four MEDIUM and one HIGH.

Figure 5 compares the signal to noise ratio (sensitivity) between the top-down science requirements and bottom-up estimates based on the subsystem conceptual designs performance. McConnachie[12] explains in the *Modeling, Systems Engineering, and Project Management for Astronomy Conference 10705* the derivations of the current sensitivity performance and outlines our mitigation plan to reconcile the top-down and bottom-up values.

The System design review panel, chaired by Michael Strauss (Princeton University), summarized their assessment of the project - "the bottom line is that this project is in very good shape, and at an appropriate level of maturity for the end of the Conceptual Design Phase. We have been very impressed by the level of sophistication that the MSE project team has brought to this project, and the tremendous amount of hard work that has been carried out thus far. This level professionalism bodes well for the project as it enters the Preliminary Design Phase."

One of the strongest recommendations from the review panel is for the project to develop a Design Reference Mission (DRM), a.k.a. the Design Reference Survey (DRS) within the MSE project. The review panel stated for a survey driven project like MSE, the DRS is an enormously useful tool for evaluating at every stage whether or not the science goals will be met by the System.

Basically, the DRS is a "narrative" document that distills the science requirements and science case into an executable survey plan for early operations, taking into account both external constraints (weather, lunar cycle, sky availability as a function of time of the year), as well as the technical constraints (observatory, instrument, and calibration) using design information collected at each design phase. In other words, the DRS represents a step-by-step plan to execute the envisaged observations, and it should inform functionally and operationally the adequacy of the observastory product breakdown structure and the PESA software tools to complete the intended observations. The DRS will also identify the precursor astronomical datasets needed to achieve the MSE specific scientific goals.

To this end, we have issued a call to the astronomical community to join the MSE internaitonal science team with a goal to "reaffirm" the science case prior to starting the DRS process. Since the call, the science team has grown to 198 astronomers, with 96 new members, at the writing of this paper. The new members of the science team will be assigned to different science working groups, led by the MSE science advisory group, according to their science interests. These range from stellar astrophysics and exoplanets to transients and time domain researc. Our plan is to kickoff the DRS in the June/July 2018 timeframe.

We realize generating scientific interest and participation in MSE are essential to build the international partnership required for the project. Through their astronomical communities, we encourage the scientists to promote and raise support for the project nationally, especially to their funding sources. The PO will continue to promote MSE's science merits in the US and internationally with a goal to engage new partners and help secure funding through our international science team and agencies.

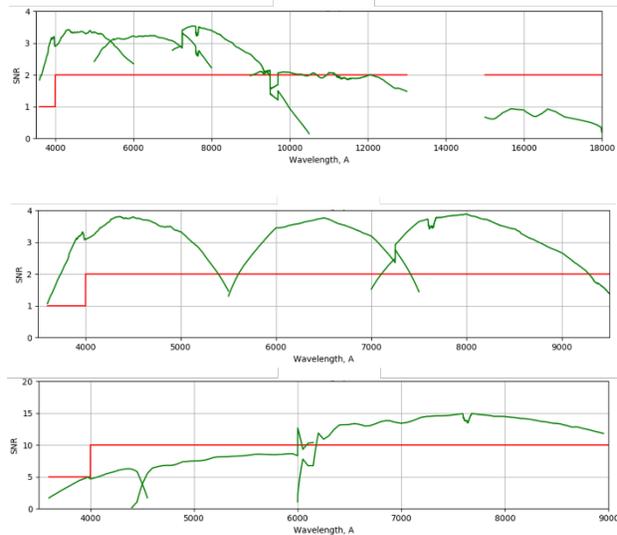

Figure 5 SNR plots - LR (top), MR (middle) and HR (bottom), science requirements (red curves) and CoDP bottom-up estimates (green curves).

## 4. PRELIMINARY DESIGN PHASE DEVELOPMENT PLAN

**4.1 Preliminary Design Phase Governance**

The planned PDP start of the MSE project is in early 2019. The current PDP design team includes members from the Australian Astronomical Observatory; National Research Council (NRC) of Canada; National Astronomical Observatories (NAOC), Chinese Academy of Sciences; Centre National de la Recherche Scientifique (CNRS) of France; University of Hawaii; India Institute of Astrophysics. They were also former participants in the CoDP.

Currently, representatives from these six organizations form the MSE management group, which is a precursor of the formal MSE Board. In addition, representatives for the National Optical Astronomy Observatory of USA and Texas A&M University are observers in the management group, exploring the possibility of joining as partners in the future. The formal governance agreement for the PDP is being developed as a Statement of Understanding (SoU) between CFHT Corp. and the MSE project participants.

The SoU empowers the management group to set the direction of the project throughout the PDP and tasks the management group to define the formal partnership in subsequent phases.

**4.2 Preliminary Design Phase Management Plan**

Similar to the CoDP, execution of the PDP work will be organized by the work breakdown structure (WBS) and formalized by work package documents between the performing organizations and the PO. Each work package document defines the scope of work, schedule (with milestones), deliverables, designated personnel and the corresponding work (to be performed) quantified in hours, and financial contributions from the performing organization and PO, such as procurements, travel costs and reimbursements. The value of each work package is the sum of the personnel costs[4] and the net financial contribution of the performing organization.

All work packages will be authorized and approved by the MSE management group based on recommendations from the PO. The values of the work packages will be accumulated toward the final share of the Observatory according to the pending MSE partnership agreement. In the same way, the accumulated CoDP contributions will be credited toward perspective partner shares.

Based on the WBS, we developed a PDP work plan outlining the work envisaged and estimations for the corresponding hours of work. Figure 6 illustrates some of the itemized Project Office costs to execute the PDP. The anticipated PDP start is in early 2019.

| WBS | Title | PDP estimate | Contingency | Total | Flag | Amount |
|---|---|---|---|---|---|---|
| MSE.PO.MGT.LOE | Management Level of Effort | $492,441 | $49,244 | $541,685 | | $541,685 |
| MSE.PO.MGT.SOFT | Software Services | $18,750 | $1,500 | $20,250 | | |
| MSE.PO.MGT.HR | Human Resources | | | | | |
| MSE.PO.MGT.PuOu | Public Outreach | $395,058 | $39,506 | $434,564 | | $434,564 |
| MSE.PO.MGT.BUS | Business Services | $255,087 | $25,508 | $280,595 | | $280,595 |
| MSE.PO.MGT.MEE | Meetings and Review | $400,000 | $0 | $400,000 | | |
| MSE.PO.MGT.PER | Permitting | $325,000 | $74,750 | $399,750 | | |
| MSE.PO.MGT.SAF | Safety | | | | | |
| MSE.PO.MGT.PA | Product Assurance | $34,011 | $3,401 | $37,412 | | |
| MSE.PO.SCI.LOE | Science Level of Effort | $863,808 | $172,761 | $1,036,569 | | $450,139 |
| MSE.PO.ENG.LOE | Engineering Level of Effort | $2,211,627 | $221,162 | $2,432,789 | | $486,686 |
| MSE.PO.ENG.STA | Engineering Standards | | | | | |
| MSE.PO.ENG.SYS | System Design | | | | | |

Figure 6 Example of PDP work plan - Project Office cost estimates

The proposed PDP work plan was circulated among the participants to solicit contributions. To date, about half of the PDP work (in value) has been contributed in-kind. All WBS elements considered science instruments[13] have been claimed, as well as leading elements in the OESA and PESA. The PO and management group continue to work together to raise the additional contributions required for the remaining elements.

## 5. PLAN FOR THE PERMITTING PROCESS

Essential for the actual construction of MSE on Maunakea will be the permits, land agreements, and community engagement needed for any new or repurposed facility on Maunakea in the future. The conflict over TMT on Maunakea is well documented[14,15] (Nature and Atlantic articles) and currently remains unresolved. Nonetheless, there are numerous important differences between building a new observatory at a new site on Maunakea and repurposing an existing observatory without changing the footprint on the ground, consistent with MSE's design (Figure 7). Cognizant of these differences, CFHT has had an important role among the Maunakea Observatories in advancing and deepening the engagement of the Observatories with the community, taking a different approach by bringing the community into our facilities through innovative programs described below.

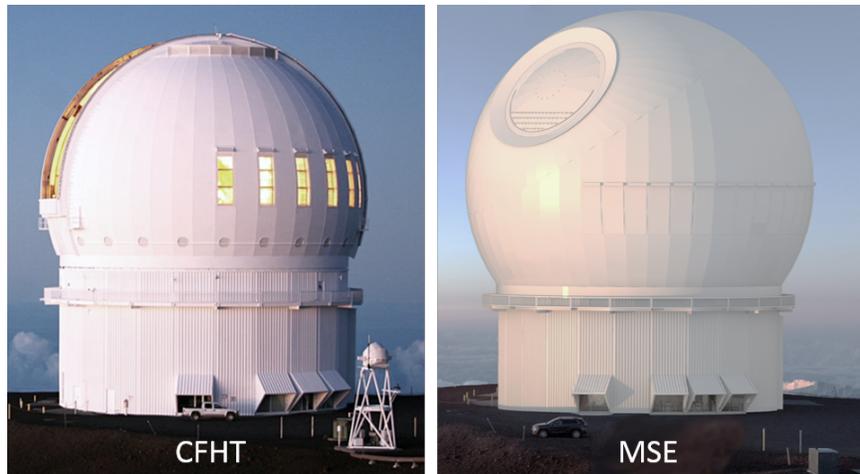

Figure 7 Exterior views of CFHT and MSE.

---

[4] Calculated using a universal labor rate, $110 USD per hour was used for the CoDP work.

The Maunakea Observatories function under a Master Lease between the State of Hawaii Department of Land and Natural Resources (DLNR) and University of Hawaii (UH). The ~11,000 acre Maunakea Science Reserve (MKSR) controlled under this Master Lease includes a ~500 acre "astronomy precinct". This is the only area in which telescopes are allowed on the summit. The MKSR also includes hundreds of culturally important sites (shrines, burial sites, etc.) surrounding the astronomy precinct that are monitored and protected by UH as part of the Master Lease arrangements. Non-UH telescopes have subleases to this Master Lease, which represent long-term commitments by the State of Hawaii to the various Maunakea Observatories to retain access to the summit and, through that, remain operational. The Master Lease expires in 2033 and if a new agreement is not reached in advance, all observatories will be removed from Maunakea by then. Thus far MSE has not pursued an EIS or the various permits from the Office of Maunakea Management (OMKM) or Department of Land and Natural Resources (DLNR) needed for construction. Instead, MSE is awaiting completion of a new long-term land authorization agreement between the State of Hawaii and UH to replace the current Master Lease. Once that agreement is in place (~2019), which should extend through most of this century, a new sub-agreement will be sought between CFHT/MSE and UH, as well as the necessary EIS and permits from DLNR and OMKM. As a result, this approach is fundamentally linked to the future of all Maunakea Observatories and is effectively decoupled from the outcome of the conflict over TMT. Combined with the fact that CFHT's footprint on the ground does not increase through the MSE conversion, and the longstanding (40 year) relationship between CFHT and the Hawaii Island community, we remain confident that once a new long term land authorization agreement is in place with UH, MSE will not suffer from the same wide spread opposition felt by TMT.

Independent of the administrative procedures used to secure a future for MSE on Maunakea, the deeper matter of nurturing the relationship between the Maunakea Observatories and the community remains central to CFHT's mission. All of the Maunakea Observatories support robust outreach programs, but CFHT has taken strides in recent years with new community engagement approaches, recognizing that resolving the conflict over Maunakea is generational in nature and requires a long-view that includes new perspectives about Maunakea from the Observatories. The reverse is also true – greater visibility by the community into the operations, research, technology and staffs of the Observatories will also help nurture a sense of community ownership in and pride about the future of Maunakea astronomy.

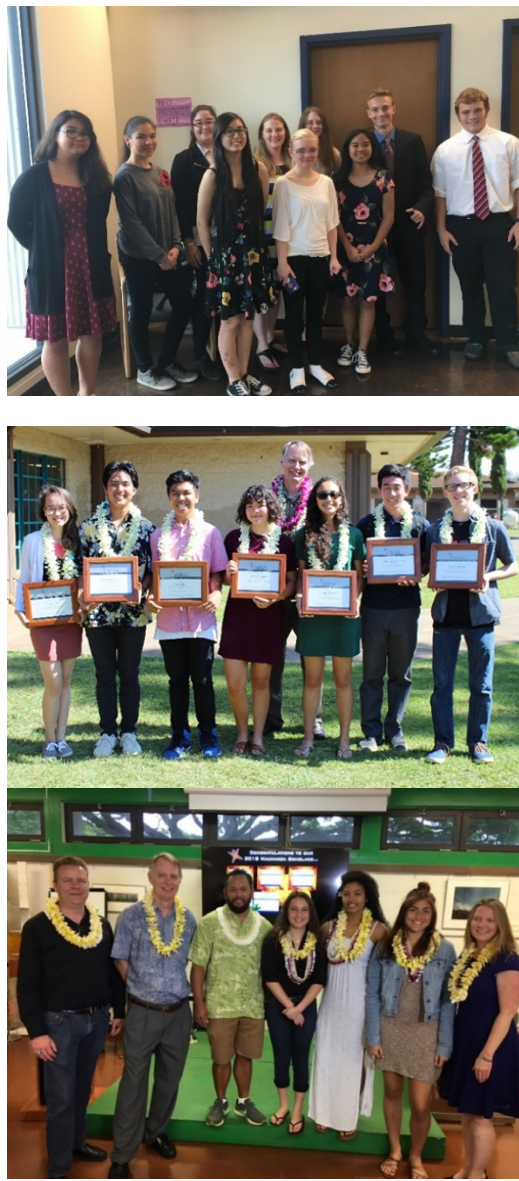

Figure 8 Students, teachers, and mentors from high schools on Hawaii Island (top), Maui (middle) and Molokai (bottom) participating in the Maunakea Scholars program are shown. With hundreds of students involved across the Hawaiian islands, this program is helping connect the Maunakea Observatories with the community at a level never before attempted.

Examples of new programs consistent with these inclusive goals include the Kamaʻāina Observatory Experience, Maunakea Fund, EnVision Maunakea, and the Maunakea Scholars program, (Figure 8). These new programs all engage the community in unique ways that have never been attempted before and, as such, are somewhat experimental. The Kamaʻāina Observatory Experience is a free monthly program available to Hawaii residents that includes presentations about the cultural importance and environmental sensitivity of the Maunakea summit, before bringing them to the summit to see two telescopes up-close. On-line registration for this program leads to all 48 openings being filled within minutes – an indication of the popularity of the program and interest in the community to see Maunakea and the observatories. The Maunakea Fund is a signature fund created within Hawaii Community Foundation's portfolio that is dedicated to sponsoring place based STEM education, environmental, and cultural programs related to Maunakea. To date $250,000 has been provided through the Maunakea Fund to sponsor a multitude of programs. EnVision Maunakea is a program that creates comfortable settings for small groups within the community to meet and discuss their visions for the future of Maunakea. After considerable preparation in 2016-2017 this program held 15 "listening sessions" or 'Aha Kūkā in different island communities. Common threads were reported to the public in a final report, the intent being that the "soft voices" captured in this report will be heard through upcoming decisions and policies pertaining to Maunakea. Finally, the Maunakea Scholars program is a partnership between the Maunakea Observatories, UH, and State Department of Education that provides mentored research opportunities for high school students across Hawaii. Students submit observing proposals which are evaluated for scientific merit and technical feasibility. Those awarded observing time are matched with one of the observatories on Maunakea. Conducted in 10 public high schools, the reaction to this program has been overwhelmingly positive. These and many other education and outreach efforts, conducted on a sustained basis, will help deepen the relationship between the Hawaii Island community and the Maunakea Observatories. By ensuring that the many interests in Maunakea are woven into a single inclusive outlook, where cultural, environmental, scientific, recreational, commercial, and other perspectives are integrated, what may be viewed as finding a balance between competing interests will actually become a foundation of common interests.

## 6. SUMMARY

The MSE project has successfully completed its CoDP. The System and its subsystem conceptual designs were reviewed by independent external panels with favorable evaluations. Incorporating the reviews recommendations, we have developed a PDP work plan, management structure (MSE management group) and scientific oversight (Design Reference Survey) to guide the project into the next phase.

Given our CoDP experience to direct and coordinate development work across a geographically distributed and culturally diverse design team, we realize the PDP will be both challenging and exciting. Resource and budget pressures will be dynamic and sometimes in short demand. This will require ongoing re-planning, anticipation and adaptability to progress the project while working with the MSE management group to maintain schedule and minimize risk exposure.

During the CoDP, the astronomical community has reaffirmed MSE as scientifically relevant, desirable and important in the era of large imaging surveys and ELT projects. We are optimistic additional resources and budget will be available as the MSE partnership grows in the coming years.

With respect to our work plan which was presented at the last SPIE conference in 2016, we have made substantial progress: The international design team has completed the enclosure and telescope structure conceptual designs, selected a multiplexing configuration, developed the LMR and HR[16,17] spectrograph designs, and established a fibre testing facility[18] to characterize the fibre transmission system performance. In conjunction, the PO has established the science calibration methodology[19], system budgets[20], and observatory operations concept[21,22]. Together, the CoDP work culminated in a successful System level conceptual design review in January 2018.

In conclusion, the project is on a positive trajectory, technically and programmatically. The project has developed a sound Observatory design with a set of enabling capabilities that are of clear interest to the astronomical community, as evidenced by recent the US and European reports[23,24] on wide field spectroscopy. We expect the MSE partnership will continue to grow and start the next design phase in 2019. We look forward to the next SPIE conference in 2020 to report on our PDP progress.

## ACKNOWLEDGEMENTS

The Maunakea Spectroscopic Explorer conceptual design phase was conducted by the MSE Project Office, which is hosted by the Canada-France-Hawaii Telescope. MSE partner organizations in Canada, France, Hawaii, Australia, China, India, and Spain all contributed to the conceptual design. The authors and the MSE collaboration recognize and acknowledge the cultural importance of the summit of Maunakea to a broad cross section of the Native Hawaiian community.